\documentclass[journal]{IEEEtran}

\usepackage{amsmath, graphicx}
\newtheorem{theorem}{\bf Theorem}
\newtheorem{conjecture}{\bf Conjecture}
\newtheorem{lemma}{\bf Lemma}
\newtheorem{corollary}{\bf Corollary}
\newtheorem{proposition}{\bf Proposition}
\newtheorem{definition}{\bf Definition}

\begin{document}
\title{Monotonic Convergence in an Information-Theoretic Law of Small Numbers}

\author{Yaming~Yu, {\it Member, IEEE}\\
\thanks{Yaming Yu is with the Department of Statistics, University of 
California, Irvine, CA, 92697-1250, USA (e-mail: yamingy@uci.edu).  This work is supported in part by a start-up fund from 
the Bren School of Information and Computer Sciences at the University of California, Irvine.}
}

\maketitle

\begin{abstract}
An ``entropy increasing to the maximum'' result analogous to the entropic central limit theorem (Barron 1986;
Artstein et al. 2004) is obtained in the discrete setting.  This involves the thinning operation and a Poisson
limit.  Monotonic convergence in relative entropy is established for general discrete distributions,
while monotonic increase of Shannon entropy is proved for the special class of ultra-log-concave distributions.  Overall we
extend the parallel between the information-theoretic central limit theorem and law of small numbers explored by Kontoyiannis
et al.\ (2005) and Harremo\"{e}s et al.\ (2007, 2008).  Ingredients in the proofs include convexity, majorization, and
stochastic orders.
\end{abstract} 


\section{Introduction}
The information-theoretic central limit theorem (CLT, \cite{B}) states that, for a sequence of independent and 
identically distributed (i.i.d.) random variables $X_i,\ i=1, 2,\ldots,$ with zero mean and unit variance, the normalized 
partial sum $Z_n=\sum_{i=1}^n X_i/\sqrt{n}$ tends to N$(0,1)$ as $n\rightarrow \infty$ in relative entropy, as long as the 
relative entropy $D(Z_n|{\rm N}(0,1))$ is eventually finite.  An interesting feature is that $D(Z_n|{\rm N}(0,1))$ decreases 
monotonically in $n$, or, equivalently, the differential entropy of $Z_n$ increases to that of the standard normal.  While 
this monotonicity is an old problem (\cite{Lieb}), its full solution is obtained only recently by Artstein et al.\ 
\cite{ABBN2}; see Tulino and Verd\'{u} \cite{TV}, Madiman and Barron \cite{MB}, and Shlyakhtenko \cite{Sh, pnas} for 
ramifications.  In this paper we establish analogous results for a general version of the law of small numbers, extending the 
parallel between the information-theoretic CLT and the information-theoretic law of small numbers explored in \cite{H} 
\cite{KHJ} \cite{HJK} and \cite{HJK2}.  Such monotonicity results are interesting as they reveal fundamental connections 
between probability, information theory, and physics (the analogy with the second law of thermodynamics).  
Moreover, the associated inequalities are often of great practical significance.  The entropic CLT, for example, is closely 
related to Shannon's entropy power inequality (\cite{Bl, S}), which is a valuable tool in analyzing Gaussian 
channels. 


Informally, the law of small numbers refers to the phenomenon that, for random variables $X_i$ on $\mathbf{Z}_+=\{0, 
1, \ldots\}$, the sum $\sum_{i=1}^n X_i$ has approximately a Poisson distribution 
with mean $\lambda=\sum_{i=1}^n EX_i$, as long as i) each of $X_i$ is such that $\Pr(X_i=0)$ is close to one, $\Pr(X_i=1)$ is 
uniformly small, and $\Pr(X_i> 1)$ is negligible compared to $\Pr(X_i=1)$; and ii) the dependence between the $X_i$'s is 
sufficiently weak.  In the version considered by Harremo\"{e}s et al.\ \cite{HJK} \cite{HJK2} and in this paper, the $X_i$'s 
are i.i.d.\ random variables obtained from a common distribution through {\it thinning}.  (Indeed, Harremo\"{e}s et al.\ 
term their result ``the law of thin numbers''.)  The notion of thinning is introduced by R\'{e}nyi \cite{R}.

\begin{definition}
The $\alpha$-thinning ($\alpha\in (0,1)$) of a probability mass function (pmf) $f$ on $\mathbf{Z}_+$, denoted as 
$T_\alpha(f)$, is the pmf of $\sum_{i=1}^Y X_i$, where $Y$ has pmf $f$ and, independent of $Y$, $X_i,\ i=1,2,\ldots$, are 
i.i.d.\ Bernoulli($\alpha$) random variables, i.e., $\Pr(X_i=1)=1-\Pr(X_i=0)=\alpha$. 
\end{definition} 

Thinning is closely associated with certain classical distributions such as the Poisson and the binomial.  For the Poisson pmf 
$po(\lambda)=\{po(i; \lambda),\ i=0,1,\ldots\}$, with $po(i; \lambda)=\lambda^i e^{-\lambda}/i!$, we have 
$$T_\alpha (po(\lambda))=po(\alpha\lambda).$$  
For the binomial pmf $bi(n,p)=\{bi(i; n,p),\ i=0,\ldots, n\}$, with $bi(i; 
n,p)=\binom{n}{i}p^i(1-p)^{n-i}$, we have 
$$T_\alpha (bi(n, p))=bi(n, \alpha p).$$  
Basic properties of thinning also include the semigroup relation (\cite{J})
\begin{equation}
\label{semi}
T_\alpha(T_\beta(f))=T_{\alpha\beta}(f).
\end{equation}
Thinning for discrete random variables is analogous to scaling for their continuous counterparts. 

The $n$-th convolution of $f$, denoted as $f^{*n}$, is the pmf of $\sum_{i=1}^n Y_i$ where $Y_i$'s are i.i.d.\ with pmf $f$.  
It is easy to show that thinning and convolution operations commute, i.e., 
\begin{equation}
\label{convcom}
T_\alpha(f^{*n})=(T_\alpha (f))^{*n}.
\end{equation}
Using the notions of thinning and convolution, we can state the following version of the law of small numbers considered by  
Harremo\"{e}s et al.\ \cite{HJK}.  As usual, for two pmfs $f$ and $g$, the entropy of $f$ is defined as $H(f)=-\sum_{i} f_i 
\log(f_i)$, and the relative entropy between $f$ and $g$ is defined as $D(f|g)=\sum_{i} f_i\log (f_i/g_i)$.  It is understood 
that $D(f|g)=\infty$ if the support of $f$, $supp(f)=\{i:\ f_i>0\}$, is not a subset of $supp(g)$.  
We frequently consider the relative entropy between a pmf $f$ and $po(\lambda)$, where $\lambda$ is the mean of $f$; we denote  
$$D(f)=D(f|po(\lambda))$$ 
for convenience. 

\begin{theorem}
\label{thm1}
Let $f$ be a pmf on $\mathbf{Z}_+$ with mean $\lambda<\infty$.  Then, as $n\to\infty$, 
\begin{enumerate}
\item
$T_{1/n}(f^{*n})$ tends to $po(\lambda)$ pointwise;
\item
$H(T_{1/n}(f^{*n}))\to H(po(\lambda))$;
\item
if $D(T_{1/n}(f^{*n}))$ ever becomes finite, then it tends to zero.
\end{enumerate}
\end{theorem} 

Part 1) of Theorem \ref{thm1} is proved by Harremo\"{e}s et al.\ \cite{HJK}, who also present a proof of Part 3) assuming 
$D(f)<\infty$.  The current, slightly more general form of Part 3) is reminiscent of Barron's work \cite{B} on the CLT.  In 
Section II we present a short proof of Part 3).  We also note that Part 2), which is stated in \cite{HJK2} with a stronger 
assumption, can be deduced from 1) directly. 

A major goal of this work is to establish monotonicity properties in Theorem \ref{thm1}.  We show that, in Part 3) of Theorem 
\ref{thm1}, the relative entropy never increases (Theorem \ref{mono}), and, assuming $f$ is {\it ultra-log-concave} (see 
Definition \ref{def2}), in Part 2) of Theorem \ref{thm1}, the entropy never decreases (Theorem \ref{thm2}).  Both Theorems 
\ref{mono} and \ref{thm2} can be regarded as discrete analogues of the monotonicity of entropy in the CLT (\cite{ABBN2}), with 
thinning playing the role of scaling.  (Unlike the CLT case, here monotonicity of the entropy and that of the relative entropy 
are not equivalent.)  We begin with monotonicity of the relative entropy. 

\begin{theorem}
\label{mono}
If $f$ is a pmf on $\mathbf{Z}_+$ with a finite mean, then $D(T_{1/n}(f^{*n}))$ decreases on $n=1, 2, \ldots$.
\end{theorem}

The proof of Theorem \ref{mono} uses two Lemmas, which are of interest by themselves.  These deal with the behavior of 
relative entropy under thinning (Lemma \ref{lemthin}) and convolution (Lemma \ref{lemconv}) respectively.  Lemma \ref{lemthin} 
is proved in Section III, where we also note its close connection with modified logarithmic Sobolev inequalities (Bobkov and 
Ledoux \cite{BL}; Wu \cite{W}) for the Poisson distribution. 

\begin{lemma}[The Thinning Lemma]
\label{lemthin}
Let $f$ be a pmf on $\mathbf{Z}_+$ with a finite mean.  Then 
$$D(T_\alpha(f))\leq \alpha D(f),\quad 0< \alpha< 1.$$
\end{lemma}

An equivalent statement is that $\alpha^{-1} D(T_\alpha(f))$ increases in $\alpha\in (0,1]$, in view of the semigroup 
property (\ref{semi}). 

Combined with a data processing argument, Lemma \ref{lemthin} can be used to show that the relative entropy is monotone 
along power-of-two iterates in Theorem \ref{mono}.  To prove Theorem \ref{mono} fully, however, we need the following 
convolution result, which may be seen as a ``strengthened data processing inequality.''  Lemma \ref{lemconv} is proved in 
Section IV. 

\begin{lemma}[The Convolution Lemma]
\label{lemconv}
If $f$ is a pmf on $\mathbf{Z}_+$ with a finite mean, then $(1/n)D\left(f^{*n}\right)$ decreases in $n$.
\end{lemma}

The main difference in the development here, compared with the CLT case, is that we need to consider the effect of both 
thinning and convolution.  In the CLT case, the monotonicity of entropy can be obtained from one general convolution 
inequality for the Fisher information (\cite{ABBN2, MB}).  Nevertheless, the proofs of Lemmas \ref{lemthin} and \ref{lemconv} 
(Lemma \ref{lemconv} in particular) somewhat parallel the CLT case.  We first express the desired divergence quantity as an
integral via a de Bruijn type identity (\cite{S, Bl, B}), and then analyze the monotonicity property of the integrand; see
Sections III and IV for details. 

Once we have Lemmas \ref{lemthin} and \ref{lemconv}, Theorem \ref{mono} is quickly established. 
\begin{IEEEproof}[Proof of Theorem \ref{mono}]
Lemma \ref{lemthin} and (\ref{semi}) imply ($n\geq 2$)
$$\frac{n}{n-1}D(T_{1/n}(f^{*n}))\leq D\left(T_{1/(n-1)}(f^{*n})\right).$$
Lemma \ref{lemconv} and (\ref{convcom}) then yield 
$$D\left(T_{1/(n-1)}(f^{*n})\right)\leq \frac{n}{n-1} D\left(T_{1/(n-1)}\left(f^{*(n-1)}\right)\right)$$
and the claim follows. 
\end{IEEEproof}

By a different analysis, we also establish the monotonicity of $H(T_{1/n}(f^{*n}))$, under the assumption that $f$ 
is ultra-log-concave.  

\begin{definition}
\label{def2}
A nonnegative sequence $u=\{u_i,\ i\in \mathbf{Z}_+\}$ is called {\it log-concave}, if the support of $u$ is an interval 
of consecutive integers, and $u_i^2\geq u_{i-1}u_{i+1}$ for all $i>0$.  A pmf $f$ is {\it ultra-log-concave}, or ULC, 
if the sequence $i!f_i,\ i\in \mathbf{Z}_+$, is log-concave. 
\end{definition}

Equivalently, $f$ is ULC if $if_i/f_{i-1}$ decreases in $i$.  It is clear that ultra-log-concavity implies log-concavity.  
Examples of ULC pmfs include the Poisson and the binomial.  More generally, the pmf of $\sum_{i=1}^n X_i$ is ULC if $X_i$'s 
are independent (not necessarily identically distributed) Bernoulli random variables.

The monotonicity of entropy is stated as follows.

\begin{theorem}
\label{thm2}
If $f$ is ULC, then $H(T_{1/n}(f^{*n}))$ increases monotonically on $n=1, 2, \ldots$.
\end{theorem}

An example (\cite{H} \cite{Yu}) is when $f$ is a Bernoulli with parameter $p$, in which case $T_{1/n}(f^{*n})=bi(n, p/n)$.  
In other words, both the entropy and the relative entropy are monotone in the classical binomial-to-Poisson convergence. 

It should not be surprising that we make the ULC assumption; the situation is similar to that of a Markov chain with 
homogeneous transition probabilities (\cite{Cover}, Chapter 4): relative entropy always decreases, but entropy does not 
increase without additional assumptions.  The ULC assumption is natural in Theorem \ref{thm2} because ULC distributions with 
the same mean $\lambda$ form a natural class in which the ${\rm Po}(\lambda)$ distribution has maximum entropy \cite{J}.  In 
fact, if we reverse the ULC assumption (but still assume that $f$ is log-concave), then $H(T_{1/n}(f^{*n}))$ decreases 
monotonically (Theorem \ref{thm3}).  Theorems \ref{thm2} and \ref{thm3} are proved in Section VI.  The starting point in 
these proofs is a general result (Lemma \ref{key}) that relates entropy comparison to comparing the expectations of convex 
functions.  This entails a rather detailed analysis of the convex order (to be defined in Section V) between the relevant 
distributions. 

As a simple example, Fig.\ 1 displays the values of 
$$\begin{array}{ll}
d(n) =D(T_{1/n}(f^{*n})),   & t(n) =nD(T_{1/n}(f)),\\
r(n) =n^{-1}D(f^{*n}),      & h(n) =H(T_{1/n}(f^{*n}))
\end{array}
$$
for $f=bi(2, 1/2)$ and $n=1, \ldots, 10$.  The monotone patterns of $d(n),\ t(n),\ r(n)$ and $h(n)$ illustrate Theorem 
\ref{mono}, Lemma \ref{lemthin}, Lemma \ref{lemconv}, and Theorem \ref{thm2} respectively. 

\begin{figure}
\begin{center}
\includegraphics[width=2.7in, height=3.4in, angle=270]{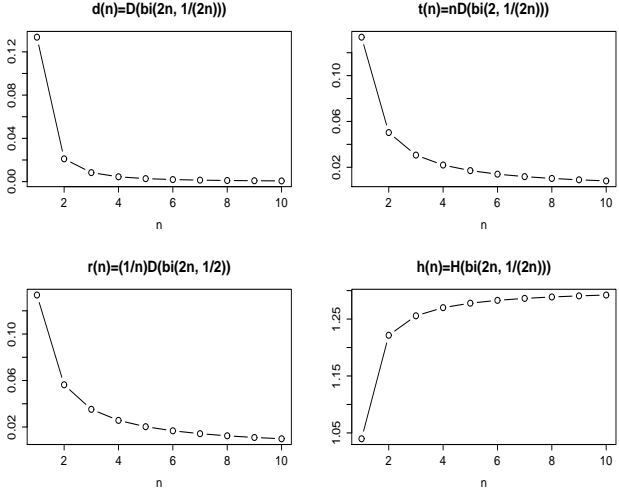}
\end{center}
\caption{Values of $d(n),\ t(n),\ r(n)$ and $h(n)$ for $n=1, \ldots, 10$.}
\end{figure}

Besides monotonicity, an equally interesting problem is the rate of convergence.  In Section VII we 
show that, if $f$ is ULC or has finite support, then $D(T_{1/n}(f^{*n}))=O(n^{-2}),\ n\to\infty$.  This complements certain 
bounds obtained by Harremo\"{e}s et al. \cite{HJK, HJK2}.  Different tools contribute to this $O(n^{-2})$ rate.  For ULC 
distributions we use stochastic orders as in Section VI; for distributions with finite support, we simply analyze the {\it 
scaled Fisher information} (\cite{KHJ, MJK}).  We conclude with a discussion on possible extensions and refinments 
(of Theorem \ref{mono} in particular) in Section VIII. 

\section{The convergence theorem}
This section deals with Theorem \ref{thm1}.  Part 1) of Theorem \ref{thm1} is proved in \cite{HJK}.  Part 2) is stated in 
\cite{HJK2} with the assumption that $f$ is ultra-logconcave.  The present form only assumes that $\lambda$, the mean of $f$, 
is finite.  Part 2) can be quickly proved as follows.  Part 1) and Fatou's lemma yield 
$$\liminf_{n\to\infty} H(T_{1/n}(f^{*n}))\geq H(po(\lambda)).$$
Let $g$ denote the pmf of a geometric($p$) distribution, i.e., $g_i=p(1-p)^i,\ i=0, 1, \ldots,\ 0<p<1$.  By the 
lower-semicontinuity property of relative entropy, 
\begin{equation}
\label{liminf}
\liminf_{n\to\infty} D(T_{1/n}(f^{*n})|g)\geq D(po(\lambda)|g).
\end{equation}
Since the mean of $T_{1/n}(f^{*n})$ is $\lambda$ for all $n$, (\ref{liminf}) simplifies to
$$\limsup_{n\to\infty} H(T_{1/n}(f^{*n}))\leq H(po(\lambda))$$
and Part 2) is proved. 

Our proof of Part 3) uses convexity arguments that also yield some interesting intermediate results (Propositions \ref{iid1} 
and \ref{iid2}).  In Propositions \ref{iid0}-- \ref{iid2} let $X_1, X_2, \ldots, $ be i.i.d.\ with pmf $f$. 
\begin{proposition}
\label{iid0}
For any $\alpha\in (0,1]$, $D(T_\alpha(f))<\infty$ if and only if $EX_1\log(X_1)<\infty$ (as usual $0\log 0=0$).
\end{proposition}
\begin{IEEEproof}
Let us consider $\alpha=1$ first.  Note that $H(f)$ is finite since the mean of $f$ is finite.  We have
\begin{equation}
\label{dh}
D(f)= \sum_{i\geq 0} f_i\log(i!) -\lambda\log(\lambda)+\lambda-H(f).
\end{equation}
Thus $D(f)<\infty$ if and only if $\sum_{i\geq 0} f_i\log(i!)$ converges, 
which, by Stirling's formula, is equivalent to $EX_1\log(X_1)<\infty$.

For general $\alpha\in (0, 1]$, let $Y|X_1\sim {\rm Bi}(X_1, \alpha)$.  By the preceding argument $D(T_\alpha(f))<\infty$ if and only if $EY\log(Y)<\infty$.  However
$$E\alpha X_1\log(\alpha X_1)\leq EY\log(Y)\leq EX_1\log(X_1)$$
where the lower bound holds by Jensen's inequality.  Thus
$EY\log(Y)<\infty$ is also equivalent to $EX_1\log(X_1)<\infty$.
\end{IEEEproof}

A consequence of Proposition \ref{iid0} is that, in Part 3), 
$$D(T_{1/n}(f^{*n}))<\infty\Longleftrightarrow E\bar{X}_n\log(\bar{X}_n)<\infty.$$  
Here and in Propositions \ref{iid1} and \ref{iid2} below, $\bar{X}_n=(1/n)\sum_{i=1}^n X_i$. 
\begin{proposition}
\label{iid1}
For $n\geq 1$, 
$$D(T_{1/n}(f^{*n}))\leq \frac{\lambda}{n}+E\bar{X}_n\log\frac{\bar{X}_n}{\lambda}.$$
\end{proposition}

\begin{IEEEproof}
We borrow an idea of \cite{HJK} used in the proof of their Proposition 8.  Letting $g=f^{*n}$, we have  
\begin{align*}
D(T_{1/n}(g)) &=D\left(\sum_{k=0}^\infty g_k bi(k, 1/n)\right)\\
                          &\leq \sum_{k=0}^\infty g_k D(bi(k, 1/n)|po(\lambda))
\end{align*}
by convexity.  However, 
\begin{align*}
D(bi(k, p)|po(\lambda)) &=D(bi(k, p))+D(po(kp)|po(\lambda))\\
                          &\leq kp^2+kp\log\frac{kp}{\lambda}-kp+\lambda
\end{align*}
where the simple bound $D(bi(k, p))\leq kp^2$ (see \cite{H} for its proof) is used in the inequality.  Thus 
\begin{align*}
D(T_{1/n}(g)) &\leq \sum_{k=0}^\infty g_k \left[\frac{k}{n^2}+\frac{k}{n}\log\frac{k}{n\lambda} 
-\frac{k}{n}+\lambda\right]\\
&=\frac{\lambda}{n}+E\bar{X}_n\log\frac{\bar{X}_n}{\lambda}
\end{align*}
as required.
\end{IEEEproof}

\begin{proposition}
\label{iid2}
Denote $l_n=E\bar{X}_n\log(\bar{X}_n/\lambda)$.  Then, as $n\uparrow\infty$, $l_n$ decreases to zero if it is finite for 
some $n$. 
\end{proposition}
\begin{IEEEproof}
By Jensen's inequality, $l_n\geq 0$.  Noting $\bar{X}_n=E\bar{X}_{n-1}|\bar{X}_{n},$
we apply Jensen's inequality again to get 
$$l_n\leq E E\left[\bar{X}_{n-1}\log (\bar{X}_{n-1}/\lambda)|\bar{X}_n\right] =l_{n-1}.$$
(Essentially we are proving $\bar{X}_n\leq_{cx} \bar{X}_{n-1}$ where $\leq_{cx}$ denotes the convex order; 
see \cite{SS}.  Section V contains a brief introduction to several stochastic orders.)  Thus $l_n\downarrow l_{\infty}$, say, 
with $l_{\infty}\geq 0$. 

We show $l_\infty=0$, assuming $l_k<\infty$ for some $k$.  By symmetry 
$l_n=E\bar{X}_k\log(\bar{X}_n/\lambda),\ n\geq k$.  We may 
use this and Jensen's inequality to obtain
\begin{align}
\nonumber
l_n &\leq E \bar{X}_k\log\frac{E\bar{X}_n|\bar{X}_k}{\lambda}\\
\label{decomp}
 &= E \bar{X}_k\log\frac{k\bar{X}_k+(n-k)\lambda}{n\lambda}.
\end{align}
However, 
$$\bar{X}_k\log\frac{k\bar{X}_k+(n-k)\lambda}{n\lambda}\leq \bar{X}_k \max\left\{0,\, 
\log\frac{\bar{X}_k}{\lambda}\right\},$$
and the right hand side has a finite expectation since $l_k<\infty$.  Letting $n\to\infty$ in 
(\ref{decomp}) and using Fatou's lemma we obtain 
$$l_\infty\leq E \bar{X}_k\log\frac{\lambda}{\lambda} =0$$
which forces $l_\infty=0$.
\end{IEEEproof}

Part 3) is then a direct consequence of Propositions \ref{iid0} -- \ref{iid2}.




\section{Lemma \ref{lemthin} and a Modified Logarithmic Sobolev Inequality}
For any pmfs $\tilde{g}$ and $g$ on $\mathbf{Z}_+$, we have 
\begin{equation}
\label{dataproc}
D(T_\alpha (\tilde{g})|T_{\alpha}(g))\leq D(\tilde{g}|g).
\end{equation}
This is a special case of a general result on the decrease of relative entropy along a Markov chain (see \cite{Cover}, 
Chapter 4).
It follows from (\ref{dataproc}) and the semigroup property (\ref{semi}) that, in the setting of Lemma \ref{lemthin}, 
$D(T_\alpha(f))$ increases in $\alpha$.  This is however not strong enough to prove Lemma \ref{lemthin} yet. 

Let us recall the {\it size-biasing} operation, which often appears in Poisson approximation problems.
\begin{definition}
For a pmf $f$ on $\mathbf{Z}_+$ with mean $\lambda>0$, the sized-biased pmf, denoted by $S(f)$, is defined on $\mathbf{Z}_+$ 
as 
$$S(f)=\{(i+1)f_{i+1}/\lambda,\ i=0, 1, \ldots\}.$$ 
\end{definition} 

The formulas $S(po(\lambda))=po(\lambda)$ and $S(bi(n, p))=bi(n-1, p)$ are readily verified.  Moreover, size-biasing and 
thinning operations commute, i.e., 
\begin{equation}
\label{stcom}
T_\alpha(S(f))=S(T_\alpha(f)).
\end{equation}
Key to the proof of Lemma \ref{lemthin} is the following identity; see Johnson \cite{J} for related calculations.
\begin{lemma}
\label{deri}
Let $f=\{f_i,\ i\geq 0\}$ be a pmf on $\mathbf{Z}_+$ with mean $\lambda\in (0, \infty)$, and assume that the support of $f$ is 
finite, i.e., there exists some $k$ such that $f_i=0$ for all $i\geq k$.  Then 
\begin{equation}
\label{hprime}
\frac{d D(T_\alpha(f))}{d\alpha}=\lambda D(T_\alpha (S(f))|T_\alpha (f)),\quad \alpha\in (0,1).
\end{equation}
\end{lemma}
\begin{IEEEproof}   
Write $g=T_\alpha(f)$ for convenience, i.e., 
$$g_i=\sum_{j\geq 0} f_j bi(i; j, \alpha).$$
By direct calculation
\begin{align*}
\frac{d D(g)}{d\alpha}=&\sum_{i\geq 0} \frac{d g_i}{d\alpha}\log\frac{g_i}{po(i; \alpha\lambda)}\\
=&\sum_{i\geq 0, j\geq 1} jf_j [bi(i-1; j-1, \alpha)-bi(i; j-1, \alpha)]\\
&\times\log\frac{g_i}{po(i; \alpha\lambda)}\\
=&\sum_{i\geq 1, j\geq 1} jf_j bi(i-1; j-1, \alpha)\\
&\times\left[\log\frac{g_i}{po(i; \alpha\lambda)}-\log\frac{g_{i-1}}{po(i-1; \alpha\lambda)}\right]\\
=&\lambda\sum_{i\geq 0} \sum_{j\geq 0} (S(f))_j bi(i; j, \alpha) \log \frac{(i+1)g_{i+1}}{\lambda g_i}\\
=&\lambda D(T_\alpha(S(f))|T_\alpha(f))
\end{align*}
where the simple identity
$$\frac{d (bi(i; n, p))}{d p} =n[bi(i-1; n-1, p)-bi(i; n-1, p)]$$
is used in the second step, and Abel's summation formula in the third.  (By convention
$bi(i; n, p)=0$ if $i<0$ or $i>n$.)  All sums are finite sums since $f$ has finite support.  
\end{IEEEproof}

{\bf Remark.}  The assumption that $f$ has finite support does not appear to impose a serious limit on the applicability 
of Lemma \ref{deri}.  Of course, it would be good to see this assumption relaxed.

\begin{IEEEproof}[Proof of Lemma \ref{lemthin}]
Let us first assume that $f$ has finite support.  Then $D(T_\alpha(f))$ is obviously continuous 
on $\alpha\in [0,1]$.  Lemma \ref{deri} and (\ref{dataproc}) show that $d D(T_\alpha(f))/d\alpha$ increases on $\alpha\in 
(0,1)$.  Thus $D(T_\alpha(f))$ is convex on $\alpha\in [0,1]$, and the claim follows.  For general $f$, we construct a sequence of pmfs 
$f^{(k)}=\{f^{(k)}_i,\ i\geq 0\},\ k=1,2,\ldots,$ by truncation.  In other words, let $f^{(k)}_i=c_kf_i,\ i=0, \ldots, k$, 
where $c_k=(\sum_{i\leq k} f_i)^{-1}$, and $f^{(k)}_i=0,\ i>k$.  Assume  
$D(f)<\infty$ without loss of generality.  Then $T_\alpha (f^{(k)})$ tends to $T_\alpha 
(f)$ pointwise as $k\to \infty$.  It is also easy to show 
$$D\left(f^{(k)}\right)\to D(f),\quad k\to\infty.$$
Thus, by the finite-support result and the lower-semi-continuity property of the relative entropy, we have 
\begin{align*}
D(T_\alpha(f)) &\leq \liminf_{k\to\infty} D\left(T_\alpha \left(f^{(k)}\right)\right)\\
&\leq \liminf_{k\to \infty} \alpha D\left(f^{(k)}\right)\\
&=\alpha D(f)
\end{align*}
as required.
\end{IEEEproof}

For two pmfs $f$ and $g$ on $\mathbf{Z}_+$ with finite means, the data-processing inequality (\cite{Cover}) gives
($*$ denotes convolution)
\begin{equation}
\label{conv0}
D(T_\alpha(f)*T_\beta(g))\leq D(T_\alpha(f))+D(T_\beta(g))
\end{equation}
where $\alpha, \beta\in [0,1]$.  By Lemma \ref{lemthin}, we have
\begin{equation}
\label{conv} 
D(T_\alpha(f)*T_\beta(g))\leq \alpha D(f)+\beta D(g).
\end{equation}
This is enough to prove Theorem \ref{mono} in the special case of power-of-two iterates, i.e.,  $D(T_{1/n}(f^{*n}))$ 
decreases on $n=2^k,\ k=0, 1, \ldots$.  To establish Theorem \ref{mono} fully, we need a 
convolution inequality stronger than (\ref{conv0}), namely Lemma \ref{lemconv}; Section IV contains the details. 

A result closely related to Lemma \ref{lemthin} is Theorem \ref{logsobo1}, which was proved by Wu (\cite{W}, 
Eqn.\ 0.6) using advanced stochastic calculus tools (see \cite{BL, C0, C} for related work).  Our proof of Theorem 
\ref{logsobo1}, based on convexity, is similar in spirit to those given by \cite{C0, C}; the use of thinning appears 
new. 

\begin{theorem}[\cite{W}]
\label{logsobo1}
For a pmf $f$ on $\mathbf{Z}_+$ with mean $\lambda\in (0, \infty)$ we have
\begin{equation}
\label{sharp}
D(f)\leq \lambda D(S(f)|f).
\end{equation}
\end{theorem}
\begin{IEEEproof}
Let us assume the support of $f$ is finite.  The convexity of $h(\alpha)=D(T_\alpha(f))$ implies $h'(\alpha)\geq 
h(\alpha)/\alpha$ for all $\alpha\in (0, 1)$.  If $D(S(f)|f)<\infty$ then $supp(f)$ is an interval of consecutive 
integers including zero.  We may let $\alpha\to 1$ and obtain 
$$\lambda D(S(f)|f) =\lim_{\alpha\uparrow 1} h'(\alpha)\geq h(1)=D(f).$$
When the support of $f$ is not finite, an argument similar to the one for Lemma \ref{lemthin} applies.
\end{IEEEproof}

Theorem \ref{logsobo1} sharpens a modified logarithmic Sobolev inequality originally obtained by Bobkov and Ledoux \cite{BL}.  
\begin{corollary}[\cite{BL}, Corollary 4]
In the setting of Theorem \ref{logsobo1}, assume that $f_i>0$ for all $i\in \mathbf{Z}_+$.  Then 
\begin{equation}
\label{logsobo2}
D(f)\leq \lambda \chi^2(S(f),\, f)
\end{equation}
where $\chi^2(S(f),\, f)=\sum_i f_i\left((S(f))_i/f_i-1\right)^2$.
\end{corollary}

The inequality (\ref{logsobo2}) follows from Theorem \ref{logsobo1} and the well-known inequality between the relative 
entropy and the $\chi^2$ distance.  For an application of (\ref{logsobo2}) to Poisson approximation bounds, see \cite{KHJ}. 

\section{Relative entropy under convolution}
This section establishes Lemma \ref{lemconv}.  The starting point is an easily verified decomposition formula (Proposition 
\ref{mix}).  Proposition \ref{mix} was used by Madiman et al.\ \cite{MJK} to derive a convolution inequality 
(\cite{MJK}, Theorem III) for the scaled Fisher information, which is $\lambda\chi^2(S(f), f)$ as in (\ref{logsobo2}).  
Here we obtain a monotonicity result (Corollary \ref{mix3}) for the relative entropy $D(S(f^{*n})|f^{*n})$, which is 
instrumental in the proof of Lemma \ref{lemconv}. 

\begin{proposition}[\cite{MJK}, Eqn.\ 14]
\label{mix}
Let $q^{(i)}$ be pmfs on $\mathbf{Z}_+$ with finite means $\lambda_i,\ i=1, \ldots, n,$ respectively ($n\geq 2$).  Define 
$q=q^{(1)}*\ldots *q^{(n)}$ and $q^{(-i)}=q^{(1)}*\ldots *q^{(i-1)}*q^{(i+1)}*\ldots* q^{(n)}$ (i.e., $q^{(i)}$ is left out), 
$i=1, \ldots, n$.  Then there holds 
$$S(q)=\sum_{i=1}^n \beta_i q^{(i)}*S\left(q^{(-i)}\right)$$ 
where $\beta_i=(1-\lambda_i/\sum_{j=1}^n \lambda_j)/(n-1)$.  (In statistical terms, we have a mixture representation of 
$S(q)$.)
\end{proposition}

\begin{proposition}
\label{mix2}
In the setting of Proposition \ref{mix} we have
\begin{align}
\label{qsq}
D(q|S(q)) &\leq \sum_{i=1}^n \beta_i D\left(q^{(-i)}|S\left(q^{(-i)}\right)\right);\\
\label{sqq}
D(S(q)|q) &\leq \sum_{i=1}^n \beta_i D\left(S\left(q^{(-i)}\right)|q^{(-i)}\right).
\end{align}
\end{proposition}
\begin{IEEEproof}
We prove (\ref{qsq}); the same argument applies to (\ref{sqq}).  By convexity, Proposition \ref{mix} yields  
$$D(q|S(q))\leq \sum_{i=1}^n \beta_i D\left(q|q^{(i)}*S\left(q^{(-i)}\right)\right).$$
However, since $q=q^{(i)}*q^{(-i)}$ for each $i$, we have 
$$D\left(q|q^{(i)}*S\left(q^{(-i)}\right)\right) \leq D\left(q^{(-i)}|S\left(q^{(-i)}\right)\right)$$
by data processing, and the claim follows. 
\end{IEEEproof} 

Corollary \ref{mix3} corresponds to the case of identical $q^{(i)}$'s in Proposition \ref{mix2}.
\begin{corollary}
\label{mix3}
For any pmf $f$ on $\mathbf{Z}_+$ with mean $\lambda\in (0,\infty)$, both $D\left(S\left(f^{*n}\right)|f^{*n}\right)$
and $D\left(f^{*n}|S\left(f^{*n}\right)\right)$ decrease in 
$n$.
\end{corollary}

\begin{IEEEproof}[Proof of Lemma \ref{lemconv}]
Let us assume that $f$ has finite support first.  We have (\ref{hprime}) in the integral form 
\begin{align}
\label{mine1}
\frac{1}{n}D(f^{*n}) &=\lambda\int_0^1 D(T_\alpha(S(f^{*n}))|T_\alpha(f^{*n}))\, d\alpha\\
\label{mine2}
&=\lambda\int_0^1 D\left(S((T_\alpha(f))^{*n})|(T_\alpha(f))^{*n}\right)\, d\alpha
\end{align}
where (\ref{mine2}) holds by the commuting relations (\ref{stcom}) and (\ref{convcom}).  By Corollary \ref{mix3}, the 
integrand in (\ref{mine2}) decreases in $n$ for each $\alpha$.  Thus $(1/n)D\left(f^{*n}\right)$ decreases in $n$ as claimed.  
For general $f$, we again use truncation.  Specifically, let $f^{(k)}$ and $c_k$ be defined as in the proof of Lemma 
\ref{lemthin}.  For $n\geq 2$ let $g=f^{*n}$, and similarly let $g^{(k)}$ denote the $n$th convolution of $f^{(k)}$.  Then 
$g^{(k)}$ tends to $g$ pointwise, and the mean of $g^{(k)}$ tends to that of $g$.  Assume $D(g)<\infty$, which amounts to 
$\sum_i g_i\log(i!)<\infty$.  The argument for Part 2) of Theorem \ref{thm1} shows 
\begin{equation}
\label{conv1}
H\left(g^{(k)}\right)\to H(g),\quad k\to\infty.
\end{equation}
We also have the simple inequality 
$g^{(k)}_i\leq c_k^n g_i$ for all $i$.  Since $c_k\to 1$ as $k\to\infty$, we may apply dominated convergence to obtain 
$$\sum_i g^{(k)}_i\log(i!)\to \sum_i g_i\log(i!),\quad k\to\infty,$$
which, taken together with (\ref{conv1}), shows
$$D\left(g^{(k)}\right)\to D(g),\quad k\to\infty.$$
The finite-support result and the lower-semicontinuity property of relative entropy then yield
$$\frac{1}{n+1}D\left(f^{*(n+1)}\right)\leq \frac{1}{n} D\left(f^{*n}\right)$$
as in the proof of Lemma \ref{lemthin}.
\end{IEEEproof} 

A generalization of Lemma \ref{lemconv} is readily obtained if we use Proposition \ref{mix2} 
rather than Corollary \ref{mix3} in the above argument. 
\begin{theorem}
\label{strong} 
In the setting of Proposition \ref{mix},
$$D(q)\leq \frac{1}{n-1}\sum_{i=1}^n D\left(q^{(-i)}\right).$$
\end{theorem} 

Theorem \ref{strong} strengthens the usual data processing inequality
$$D(q )\leq \sum_{i=1}^n D\left(q^{(i)}\right)$$
in the same way that the entropy power inequality of Artstein et al.\ \cite{ABBN2} strengthens Shannon's classical entropy 
power inequality. 


{\bf Remark.} A by-product of Corollary \ref{mix3} is that the divergence quantities 
\begin{align*}
h_n &=D(T_{1/n}(f^{*n})|S(T_{1/n}(f^{*n})))\quad {\rm and} \\
\tilde{h}_n &=D\left(S(T_{1/n}(f^{*n}))|T_{1/n}(f^{*n})\right)
\end{align*}
also decrease in $n$.  Indeed we have 
\begin{align}
\nonumber
h_n &=D(T_{1/n}(f^{*n})|T_{1/n}(S(f^{*n})))\\
\label{step2}
    &\leq D(T_{1/(n-1)}(f^{*n})|T_{1/(n-1)}(S(f^{*n})))\\
\nonumber
    &= D((T_{1/(n-1)}(f))^{*n}|S((T_{1/(n-1)}(f))^{*n}))\\
    \label{step4}
    &\leq h_{n-1}
\end{align}
where (\ref{dataproc}) is used in (\ref{step2}), Corollary \ref{mix3} is used in (\ref{step4}), and the commuting 
relations (\ref{stcom}) and (\ref{convcom}) are applied throughout.  The proof for $\tilde{h}_n$ is the same.  These 
monotonicity statements complement Theorem \ref{mono}. 

\section{Stochastic orders and majorization}
The proof of the monotonicity of entropy (Theorem \ref{thm2}) involves several notions of stochastic orders which we briefly 
introduce.

\begin{definition}
For two random variables $X$ and $Y$ with pmfs $f$ and $g$ respectively, 
\begin{itemize}
\item
$X$ is smaller than $Y$ in the {\it usual stochastic order}, written as $X\leq_{st} Y$, if $\Pr(X> c)\leq \Pr(Y> c)$ 
for all $c$; 
\item
$X$ is smaller than $Y$ in the {\it convex order}, written as $X\leq_{cx} Y$, if $E \phi(X)\leq E \phi(Y)$ for every 
convex function $\phi$ such that the expectations exist; 
\item
$X$ is log-concave relative to $Y$, written as $X\leq_{lc} Y$, if i) both $supp(f)$ and $supp(g)$ are intervals of 
consecutive integers, ii) $supp(f)\subset supp(g)$, and iii) $\log(f_i/g_i)$ is concave on $supp(f)$.
\end{itemize}
\end{definition} 

We use $\leq_{st},\ \leq_{cx},\ \leq_{lc}$ with the pmfs as well as the random variables.  In general, $f\leq_{st} g$ if there 
exist random variables $X$ and $Y$ with pmfs $f$ and $g$ respectively such that $X\leq Y$ almost surely.  Examples include  
$$bi(n, p)\leq_{st} bi(n+1, p),\quad bi(n, p)\leq_{st} bi(n, p'),\ p\leq p'.$$
In contrast, $\leq_{cx}$ compares variability.  A classical example (Hoeffding \cite{H56}) is  
$$bi\left(n,\,\lambda/n\right) \leq_{cx} bi\left(n+1,\, \lambda/(n+1)\right),\quad 0\leq \lambda\leq n.$$  
Another example mentioned in Section II is 
$\bar{X}_n\leq_{cx} \bar{X}_{n-1}$ where $\bar{X}_n=(1/n)\sum_{i=1}^n X_i$ for i.i.d.\ $X_i$'s with a finite mean.
The log-concavity order $\leq_{lc}$ is also useful in our context; for example, $f$ 
being ULC can be written as $f\leq_{lc} po(\lambda),\ \lambda>0$. (The actual value of $\lambda$ is irrelevant.)  Further 
properties of these stochastic orders can be found in Shaked and Shanthikumar \cite{SS}. 

We also need the concepts of {\it majorization} and {\it Schur concavity}. 

\begin{definition}
A real vector $\mathbf{b}=(b_1, \ldots, b_n)$ is said to majorize $\mathbf{a}=(a_1, \ldots, a_n)$, written as $\mathbf{a}\prec
\mathbf{b}$, if
\begin{itemize}
\item
$\sum_{i=1}^n a_i=\sum_{i=1}^n b_i$, and
\item
$\sum_{i=k}^n a_{(i)}\leq \sum_{i=k}^n b_{(i)},\ k=2, \ldots, n,$ where $a_{(1)}\leq \ldots\leq a_{(n)}$
and $b_{(1)}\leq \ldots\leq b_{(n)}$ are $(a_1, \ldots, a_n)$ and $(b_1, \ldots, b_n)$ arranged in increasing order,
respectively.
\end{itemize}
A function $\phi(\mathbf{a})$ symmetric in the coordinates of $\mathbf{a}=(a_1, \ldots, a_n)$ is said to be {\it Schur
concave}, if
$$\mathbf{a}\prec \mathbf{b}\Longrightarrow \phi(\mathbf{a})\geq \phi(\mathbf{b}).$$
\end{definition} 

As is well-known, if pmfs $f$ and $g$ on $\{0, \ldots, n\}$ (viewed as vectors of the respective probabilities) satisfy 
$f\prec g$, then $H(f)\geq H(g)$.  In other words $H(f)$ is a Schur concave function of $f$.  Further properties and various 
applications of these two notions can be found in Hardy et al.\ \cite{HLP} and Marshall and Olkin \cite{MO}. 

\section{Monotonicity of the entropy}
This section proves Theorem \ref{thm2}.  We state a key lemma that can be traced back to Karlin and Rinott \cite{KR}. 

\begin{lemma}
\label{key}
Let $f$ and $g$ be pmfs on $\mathbf{Z}_+$ such that $f\leq_{cx} g$ and $g$ is log-concave.  Then 
$$H(f)+D(f|g)\leq H(g).$$
In particular $H(f)\leq H(g)$ with equality only if $f=g$.
\end{lemma}

Although Lemma \ref{key} follows almost immediately from the definitions (hence the proof is omitted), it is a useful tool in 
several entropy comparison contexts (\cite{KR, Yu, Yu3, Yu4}).  Effectively, Lemma \ref{key} reduces entropy comparison to two (often 
easier) problems: i) establishing a log-concavity result, and ii) comparing the expectations of convex functions.  A 
modification of Lemma \ref{key} is used by \cite{Yu} to give a short and unified proof of the main theorems of \cite{J} and 
\cite{Yu2} concerning the maximum entropy properties of the Poisson and binomial distributions.  We quote Johnson's result.  
Further extensions to compound distributions can be found in \cite{JKM, Yu4}. 

\begin{theorem}
\label{johnson}
If a pmf $f$ on $\mathbf{Z}_+$ is ULC with mean $\lambda$, then $H(f)\leq H(po(\lambda))$, with equality only if 
$f=po(\lambda)$.
\end{theorem}

To apply Lemma \ref{key} to our problem, we show that, in the setting of Theorem \ref{thm2}, 
\begin{equation}
\label{cx}
T_{1/(n-1)}(f^{*(n-1)}) \leq_{cx} T_{1/n}(f^{*n}).
\end{equation}
In a sense, (\ref{cx}) means that $T_{1/n}(f^{*n})$ becomes more and more ``spread out'' as $n$ increases.  On the other 
hand, it can be shown that $T_{1/n}(f^{*n})$ is log-concave for all $n$.  Indeed, $f$ is ULC 
and hence log-concave.  It is well-known that convolution preserves log-concavity.  That thinning preserves log-concavity is 
sometimes known as Brenti's criterion \cite{Bren} in the combinatorics literature.  Thus $T_{1/n}(f^{*n})$ remains 
log-concave.  Actually, since $f$ is ULC, there holds the stronger relation 
\begin{equation}
\label{close}
T_{1/n}(f^{*n})\leq_{lc} po(\lambda).
\end{equation}
Relation (\ref{close}) follows from i) if $f$ is ULC then so is $f^{*n}$ (Liggett \cite{L}) and ii) if $f$ is ULC then 
so is $T_\alpha(f)$ (Johnson \cite{J}, Proposition 3.7). 

The core of the proof of Theorem \ref{thm2} is proving (\ref{cx}).  The notions of majorization and Schur concavity briefly 
reviewed in Section V are helpful in formulating a more general (and easier to handle) version of (\ref{cx}).  

\begin{proposition}
\label{main}
Let $Y_1,\ldots, Y_n$ be i.i.d.\ random variables on $\mathbf{Z}_+$ with an ultra-log-concave pmf $f$.  Conditional on the 
$Y_i$'s, let $Z_i,\ i=1,\ldots, n,$ be independent ${\rm Bi}(Y_i, p_i)$ random variables respectively, where $p_1,\ldots, 
p_n\in [0,1]$.  Let $\phi$ be a convex function on $\mathbf{Z}_+$.  Then $E\phi(\sum_{i=1}^n Z_i)$ is a Schur concave function 
of $(p_1,\ldots, p_n)$ on $[0,1]^n$. 
\end{proposition}

The proof of Proposition \ref{main}, somewhat technical, is collected in the appendix.

\begin{IEEEproof}[Proof of (\ref{cx})]
Noting that 
$$(1/n, \ldots, 1/n)\prec (1/(n-1),\ldots, 1/(n-1), 0)$$
the claim follows from Proposition \ref{main} and the definition of Schur-concavity.
\end{IEEEproof} 

Theorem \ref{thm2} then follows from (\ref{cx}), (\ref{close}) and Lemma \ref{key}.  

{\bf Remark.} Theorem \ref{thm2} resembles the semigroup argument of Johnson \cite{J} in that both are statements of ``entropy 
increasing to the maximum,'' and both involve convolution and thinning operations.  The difference is that \cite{J} considers 
convolution with a Poisson while we study the self-convolution $f^{*n}$. 

As mentioned in Section I, if we reverse the ULC assumption (but still assume log-concavity), then the conclusion of Theorem 
\ref{thm2} is also reversed. 

\begin{theorem}
\label{thm3} 
Let $f$ be a pmf on $\mathbf{Z}_+$ with mean $\lambda$. Assume $f$ is log-concave, and assume $po(\lambda)\leq_{lc} f$.  Then 
$H(T_{1/n}(f^{*n}))$ decreases in $n$. 
\end{theorem} 

Theorem \ref{thm3} extends a minimum entropy result that parallels Theorem 
\ref{johnson}.

\begin{proposition}[\cite{Yu}]
\label{lculc}
The ${\rm Po}(\lambda)$ distribution achieves minimum entropy among all pmfs $f$ with mean $\lambda$ such that $f$ is 
log-concave and $po(\lambda)\leq_{lc} f$.
\end{proposition}

An example of Theorem \ref{thm3}, also noted in \cite{Yu}, is when $f$ is a geometric$(p)$ pmf, in which
case $T_{1/n}(f^{*n})=nb(n, n/(n-1+1/p))$.  (Here $nb(n, p)$ denotes the negative binomial pmf 
with parameters $(n,p)$, i.e., $nb(n,p)=\{\binom{n+i-1}{i}p^n(1-p)^{i},\ i=0,1,\ldots\}$.)  In other words, the
negative-binomial-to-Poisson convergence is monotone in entropy (as long as the first parameter of the negative binomial is 
at least 1). 

The proof of Theorem \ref{thm3} parallels that of Theorem \ref{thm2}.  In place of (\ref{cx}) we have 
\begin{equation}
\label{cx1}
T_{1/n}(f^{*n})\leq_{cx} T_{1/(n-1)}(f^{*(n-1)})
\end{equation}
assuming $po(\lambda)\leq_{lc} f$.  The proof of (\ref{cx}) applies after reversing the direction of $\leq_{lc}$ in the 
relevant places.  As noted before, since $f$ is log-concave, $T_{1/n}(f^{*n})$ is log-concave for all $n$.  Thus Theorem 
\ref{thm3} follows from Lemma \ref{key} as does Theorem \ref{thm2}. 

Incidentally, we have
\begin{equation}
\label{close1}   
po(\lambda)\leq_{lc} f \Longrightarrow po(\lambda)\leq_{lc} T_{1/n}(f^{*n}),
\end{equation}
which is a reversal of (\ref{close}).  To prove (\ref{close1}), we note that, according to a result of Davenport 
and P\'{o}lya \cite{DP}, $po(\lambda)\leq_{lc} f$ implies $po(\lambda)\leq_{lc} f^{*n}$.  By a slight modification of the argument of Johnson (\cite{J}, Proposition 3.7), we can also show that $po(\lambda)\leq_{lc} f$ implies $po(\lambda)\leq_{lc} T_\alpha (f)$ (details omitted); thus (\ref{close1}) holds. 

\section{Rate of convergence}
Assuming that $f$ is a pmf on $\mathbf{Z}_+$ with mean $\lambda$ and variance $\sigma^2<\infty$, Harremo\"{e}s et al.\ 
(\cite{HJK}, Corollary 9) show that 
$$D(T_{1/n}(f^{*n}))\leq \frac{\lambda}{n}+\frac{\sigma^2}{n\lambda}.$$
That is, the relative entropy converges at a rate of (at least) $O(n^{-1})$.  We aim to improve this to 
$O(n^{-2})$ under some natural assumptions.  The $O(n^{-2})$ rate is perhaps not surprising since, in the binomial 
case (\cite{HR}), 
\begin{equation}
\label{r2}
D(bi(n, \lambda/n))=O(n^{-2}),\quad n\to\infty.
\end{equation}

We first use the stochastic orders $\leq_{cx}$ and $\leq_{lc}$ to extend (\ref{r2}) to ULC distributions. 

\begin{theorem}
\label{rulc}
If $f$ is ULC on $\mathbf{Z}_+$ with mean $\lambda$, then  
\begin{align}
\nonumber
D(T_{1/n}(f^{*n}))\leq &\left\{n\lambda\right\}D(bi(\lfloor n\lambda\rfloor+1, 1/n)|po(\lambda))\\
\label{eq:rulc}
&+(1-\left\{n\lambda\right\}) D(bi(\lfloor n\lambda\rfloor, 1/n)|po(\lambda))
\end{align}
where $\left\{x \right\}$ and $\lfloor x\rfloor$ denote the fractional and integer parts of $x$, respectively. 
\end{theorem}

Theorem \ref{rulc} and (\ref{r2}) easily yield 
$$D(T_{1/n}(f^{*n}))=O(n^{-2}),\quad n\to\infty,$$
as long as $f$ is ULC.  To prove Theorem \ref{rulc}, we again adopt the strategy of Section VI.  Proposition \ref{key2} is a 
variant of Lemma \ref{key}. 

\begin{proposition}
\label{key2}
Let $f$ and $g$ be pmfs on $\mathbf{Z}_+$ such that $f\leq_{cx} g$ and $g$ is ULC.  Then 
$$D(f)\geq D(g)+D(f|g).$$
\end{proposition}

We also have the following result, which is easily deduced from Theorem 3.A.13 of Shaked and Shanthikumar \cite{SS} (see also \cite{Yu4}, Lemma 2).  Plainly, it says that the convex order $\leq_{cx}$ is preserved under thinning. 

\begin{proposition}
\label{cxthin}
If $f$ and $g$ are pmfs on $\mathbf{Z}_+$ such that $f\leq_{cx} g$, then $T_\alpha f\leq_{cx} T_\alpha g,\ \alpha\in (0,1)$.
\end{proposition}

\begin{IEEEproof}[Proof of Theorem \ref{rulc}] 
Let $g$ be the two-point pmf that assigns probability $\left\{ n\lambda \right\}$ 
to $\lfloor n\lambda\rfloor +1$ and the remaining probability to $\lfloor n\lambda\rfloor$.  Note that the mean of $g$ is 
$n\lambda$.  Also, the relation $g\leq_{cx} f^{*n}$ is intuitive and easily proven.  Indeed, if $\phi$ is a 
convex function on $\mathbf{Z}_+$, then 
$$\phi(x)\geq (x-\lfloor n\lambda \rfloor) \phi(\lfloor n\lambda \rfloor+1) + (\lfloor n\lambda \rfloor+1-x) \phi(\lfloor 
n\lambda \rfloor).$$
The claim follows by taking the weighted average with respect to $f^{*n}$.  By Proposition \ref{cxthin}, $T_{1/n} g \leq_{cx} 
T_{1/n} (f^{*n})$.  Since $f$ is ULC, so is $T_{1/n}(f^{*n})$.  By Proposition \ref{key2}, $D(T_{1/n}(f^{*n}))\leq D(T_{1/n} 
g)$.  However $T_{1/n} g$ is a mixture of two binomials: 
$$T_{1/n} g = \left\{ n\lambda \right\} bi(\lfloor n\lambda\rfloor +1, 1/n) +(1-\left\{ n\lambda \right\}) bi(\lfloor 
n\lambda\rfloor, 1/n).$$
Thus (\ref{eq:rulc}) holds by the convexity of the relative entropy.
\end{IEEEproof}

Although (\ref{eq:rulc}) implies the right order of the convergence rate, the bound itself does not involve the 
variance of $f$.  It is known that, if $f$ is ULC, then its variance $\sigma^2$ does not exceed its mean $\lambda$ (\cite{J, 
Yu}).  It is intuitively reasonable that the closer $\sigma^2$ is to $\lambda$, the smaller $D(f)$ and  $D(T_{1/n}(f^{*n}))$ 
are.  Hence any bound that accounts for the variance $\sigma^2$ would be interesting.  

Of course, it would also be interesting to see the ULC assumption relaxed.  Theorem \ref{rfinite} shows that the  $O(n^{-2})$ 
rate holds under a finite support assumption.  Note that, in the CLT case, an $O(n^{-1})$ rate of convergence for the 
relative entropy can be obtained under a ``spectral gap'' assumption (\cite{ABBN1, JB}); possibly a similar assumption 
suffices in our case.  Under the finite support assumption, however, the proof of Theorem \ref{rfinite} is elementary, 
although it does use a nontrivial subadditivity property of the scaled Fisher information (\cite{KHJ, MJK}). 

\begin{theorem}
\label{rfinite}
Suppose $f$ is a pmf on $\mathbf{Z}_+$ with finite support and denote the mean and variance of $f$ by $\lambda$ and $\sigma^2$ 
respectively.  Then 
\begin{equation}
\label{O-2}
D(T_{1/n}(f^{*n}))=O(n^{-2}),\quad n\to\infty.
\end{equation}
If $\lambda=\sigma^2$ in addition, then the right hand side of (\ref{O-2}) can be replaced by $O(n^{-3})$.
\end{theorem}
\begin{IEEEproof} Let us assume $\lambda>0$ to eliminate the trivial case.  For a pmf $g$ on $\mathbf{Z}_+$ with mean 
$\mu>0$, define $K(g)=\mu \chi^2(S(g), g)$ as in (\ref{logsobo2}).  Madiman et al.\ (\cite{MJK}, Theorem III) show that 
$K(g^{*n})$ decreases in $n$.  In particular, letting $g=T_{1/n}(f)$, and noting (\ref{logsobo2}) and (\ref{convcom}), we 
obtain 
$$D(T_{1/n}(f^{*n}))\leq K(T_{1/n}(f^{*n}))\leq K(T_{1/n}(f)).$$
Thus, to prove (\ref{O-2}), we only need $K(T_{1/n}(f))=O(n^{-2})$.  By the definition of $K(\cdot)$ and (\ref{stcom}), this 
is equivalent to 
\begin{equation}
\label{chi2small}
\chi^2(T_{1/n}(S(f)), T_{1/n}(f))=O(n^{-1}).
\end{equation}
However, for each $i\geq 0$ we have 
\begin{align*}
(T_{1/n}(f))_i &=\sum_{j=i}^k f_j bi(i; j, 1/n)\\
               &=n^{-i}\sum_{j=i}^k \binom{j}{i} f_j+O(n^{-i-1})
\end{align*}
where $k$ is the largest integer such that $f_k\neq 0$; a similar expression holds for $T_{1/n}(S(f))$.  
By direct calculation, each term in the sum
\begin{equation}
\label{chi2sum}
\sum_{i=0}^k \frac{((T_{1/n}(S(f)))_i - (T_{1/n}(f))_i) ^2}{(T_{1/n}(f))_i}
\end{equation}
is $O(n^{-1})$, and (\ref{chi2small}) holds.  If $\lambda=\sigma^2$, then each term in (\ref{chi2sum}) is $O(n^{-2})$, thus 
proving the remaining claim.
\end{IEEEproof}


Theorems \ref{rulc} and \ref{rfinite} imply a corresponding rate of convergence for the total variation distance, 
which is defined as $V(g, \tilde{g})=\sum_i |g_i-\tilde{g}_i|$ for any pmfs $g$ and $\tilde{g}$.  The total 
variation is related to the relative entropy via Pinsker's inequality $V^2(g, \tilde{g})\leq 2D(g|\tilde{g})$.  
Hence, if $f$ is either ULC or has finite support, then 
$$V(T_{1/n}(f^{*n}), po(\lambda))=O(n^{-1}).$$
An explicit upper bound, possibly via the Stein-Chen method, is of course desirable. 

\section{Summary and possible extensions}
We have extended the monotonicity of entropy in the central limit theorem to a version of the law of small numbers, which 
involves the thinning operation (the discrete analogue of scaling), and a Poisson limit (the discrete counterpart of the 
normal).  For a pmf $f$ on $\mathbf{Z}_+$ with mean $\lambda$, we show that the relative entropy  
$D(T_{1/n}(f^{*n})|po(\lambda))$ decreases monotonically in $n$ (Theorem \ref{mono}), and, if $f$ is ultra-log-concave, the 
entropy $H(T_{1/n}(f^{*n}))$ increases in $n$ (Theorem \ref{thm2}).  In the process of establishing Theorem \ref{mono}, 
inequalities are obtained for the relative entropy under thinning and convolution, and connections are made with logarithmic 
Sobolev inequalities and with the recent results of Kontoyiannis et al.\ \cite{KHJ} and Madiman et al.\ \cite{MJK}.  Theorem 
\ref{thm2}, in contrast, is established by comparing pmfs with respect to the convex order, an idea that dates back to Karlin 
and Rinott \cite{KR}. 


This work is arguably more qualitative than quantitative, given its focus on monotonicity.  When bounds are occasionally
obtained, in Proposition \ref{iid1} for example, we do not claim that they are always sharp.  Among the large literature on
Poisson approximation bounds (e.g., Barbour et al.\ \cite{BHJ}), the use of information theoretic ideas is a relatively new   
development (\cite{KHJ, MJK}).  We have, however, obtained an upper bound and identified an $O(n^{-2})$ rate for the relative
entropy under certain simple conditions.  Such results complement those of \cite{HJK} and \cite{HJK2}. 

The analogy with the CLT leads to further questions.  For example, given the intimate connection
between the information-theoretic CLT with Shannon's entropy power inequality (EPI), it is natural 
to ask whether there exists a discrete version of the EPI.  By analogy with the CLT, our results seem 
to suggest that the answer is yes, although there is still much to be done.  Certain simple formulations 
of the EPI do not hold in the discrete setting; see \cite{YJ} for recent developments.


We may also consider extending our monotonicity results to compound Poisson limit theorems.  Recently, Johnson et al. \cite{JKM} (see also \cite{Yu4}) have shown that compound Poisson distributions admit a maximum entropy characterization similar to that of the Poisson.  Such results suggest the possibility of compound Poisson limit theorems with the same appealing ``entropy increasing to the maximum'' interpretation.

Finally, on a more technical note, we point out a possible refinement of Theorem \ref{mono}.  This is 
analogous to the results of Yu \cite{Yu5}, who noted that relative entropy is {\it completely monotonic} in the CLT for 
certain distribution families.  (A function is completely monotonic if its derivatives of all orders exist and alternate in 
sign; the definition is similar for discrete sequences; see Feller \cite{F} for the precise statements.) 

\begin{theorem}[\cite{Yu5}]
\label{invg}
Let $X_i,\ i=1, 2,\ldots,$ be i.i.d. random variables with distribution $F$, mean $\mu$, and variance $\sigma^2\in
(0,\infty)$.  Then $D\left(\sum_{i=1}^n (X_i-\mu)/\sqrt{n\sigma^2}|{\rm N}(0,1)\right)$ is a completely monotonic 
function of $n$ if $F$ is either a gamma distribution or an inverse Gaussian distribution.
\end{theorem} 

Part of the reason that the gamma and inverse Gaussian distributions are considered is that they are analytically tractable.  
The result may conceivably hold for a wide class of distributions.  We conclude with a discrete analogue based on numerical 
evidence. 

\begin{conjecture}
Let $\lambda>0$.  Then 
\begin{itemize}
\item
$D(bi(n, \lambda/n))$ is completely monotonic in $n$ ($n\geq \lambda$);
\item
$D(nb(n, n/(\lambda+n)))$ is completely monotonic in $n$ ($n>0$).
\end{itemize}
\end{conjecture}

We again expect similar results for other pmfs, but are unable to prove even those for the binomial and the negative 
binomial. 

\appendix
\section*{Proof of Proposition \ref{main}}
Let us recall a well-known characterization of the convex order (see \cite{SS}, Theorem 3.A.1, for example).

\begin{proposition}
\label{prop1} 
Let $X$ and $Y$ be random variables on $\mathbf{Z}_+$ such that $EX=EY<\infty$.  Then
 $X\leq_{cx} Y$ if and only if
 $$E\max\{X-k,\, 0\}\leq E\max\{Y-k,\, 0\},\quad k\geq 0,$$
 or, equivalently,
 $$\sum_{i\geq k} \Pr(X\geq i)\leq \sum_{i\geq k} \Pr(Y\geq i),\quad k\geq 0.$$
\end{proposition}

\begin{proposition}
\label{prop3}
Fix $p\in (0,1)$, and let $Y_1$ and $Y_2$ be i.i.d.\ random variables on $\mathbf{Z}_+$ with an ultra-log-concave pmf
$f$.  Let $Z_1,\ Z_2,\ Z_1'$ and $Z_2'$ be independent conditional on $Y_1$ and $Y_2$ and satisfy 
$$\begin{array}{ll}
Z_1|Y_1\sim {\rm Bi}(Y_1, p+\delta), & Z_2|Y_2\sim {\rm Bi}(Y_2, p-\delta),\\
Z_1'|Y_1\sim {\rm Bi}(Y_1, p+\delta'), & Z_2'|Y_2\sim {\rm Bi}(Y_2, p-\delta').  
\end{array}
$$
If $\delta>\delta'\geq 0$, then $Z_1+Z_2\leq_{cx} Z_1'+Z_2'$.
\end{proposition}

\begin{IEEEproof}
We show that, for each $k\geq 0$, $\sum_{i\geq k} \Pr(Z_1+Z_2\geq i)$ is a decreasing function of $\delta$ as long as $0\leq 
\delta\leq \min\{p,\, 1-p\}$.  The claim then follows from Proposition \ref{prop1} (the assumptions imply 
$E(Z_1+Z_2)=E(Z_1'+Z_2')<\infty$).  To simply the notation, in what follows the limits of summation, if not spelled out, are 
from $-\infty$ to $\infty$; also $f_i\equiv 0$ if $i<0$.  Denoting $B(i; n, p)=\sum_{j\geq i} bi(j; n, p)$, and letting 
$h(\delta)=\sum_{i\geq k} \Pr(Z_1+Z_2\geq i)$, we have
\begin{align}
\nonumber
h(\delta) = & \sum_{i\geq k}\sum_j \Pr(Z_1\geq j)\Pr(Z_2= i-j)\\
\nonumber
= &\sum_j \Pr(Z_1\geq j)\Pr(Z_2\geq k-j)\\
\nonumber
=&\sum_j \left[\sum_{s\geq 0} f_s B(j; s, p+\delta)\right]\left[\sum_{s\geq 0} f_s B(k-j; s, p-\delta)\right]\\
\label{vsum}
=&\sum_{s, t\geq 0} f_s f_t v(s, t, \delta)
\end{align}
where 
$$v(s, t, \delta)=\sum_j B(j; s, p+\delta) B(k-j; t, p-\delta).$$
Using the simple identity
$$\frac{d B(i; n, p)}{d p}    = n[bi(i-1; n-1, p)]$$
we get 
$$\frac{d v(s, t, \delta)}{d \delta} = s u(s, t, \delta) - t u(t, s, -\delta)$$
where 
\begin{align*}
u(s, t, \delta)&=\sum_j bi(j-1; s-1, p+\delta)B(k-j; t, p-\delta)\\
               &=\sum_j bi(k-j-1; s-1, p+\delta)B(j; t, p-\delta).
\end{align*}
The quantity $u(s, t, \delta)$ has the following interpretation.  If we let $V_1\sim {\rm Bi}(s-1, p+\delta)$ and $V_2\sim 
{\rm Bi}(t, p-\delta)$ independently, then $u(s, t, \delta)=\Pr(V_1+V_2\geq k-1).$  Clearly 
\begin{equation}
\label{ustd}
u(s, t, \delta)=u(t+1, s-1, -\delta).
\end{equation}
Hence, we may take the derivative under the summation in (\ref{vsum}) (by dominated convergence), and then apply (\ref{ustd}) 
to obtain
\begin{align}
\nonumber
\frac{d h(\delta)}{d \delta} = &\sum_{s,t\geq 0} sf_sf_t u(s, t,\delta)-\sum_{s,t\geq 0} tf_sf_t u(t, s, -\delta)\\
\nonumber
    =&\sum_{s\geq 1,t\geq 0} sf_sf_t u(s, t,\delta)\\
    \nonumber
     &-\sum_{s\geq 1,t\geq 0} (t+1)f_{s-1}f_{t+1} u(t+1, s-1, -\delta)\\
\label{ust}
    =&\sum_{s\geq 1,t\geq 0} [sf_sf_t-(t+1)f_{s-1}f_{t+1}] u(s, t,\delta).
\end{align}
By a change of variables $s\to t+1$ and $t\to s-1$ in (\ref{ust}), and by (\ref{ustd}), we get
\begin{equation}
\label{ust2}
\frac{d h(\delta)}{d \delta}= \sum_{s\geq 1, t\geq 0} [(t+1)f_{t+1}f_{s-1} -sf_sf_t] u(s, t, -\delta).
\end{equation}
Combining (\ref{ust}) and (\ref{ust2}), and noting the symmetry, we obtain
\begin{equation}
\label{ust3}
\frac{d h(\delta)}{d \delta} = \sum_{1\leq s\leq  t} [sf_sf_t - (t+1)f_{t+1} f_{s-1}] [u(s, t,\delta)-u(s, t, -\delta)].
\end{equation}
Because $f$ is ULC, if $s\leq t$, then
\begin{equation}
\label{fst}
sf_sf_t\geq (t+1)f_{t+1}f_{s-1}.
\end{equation}
We can also show ($s\leq t$)
\begin{equation}
\label{uleq}
u(s,t, \delta)\leq u(s, t, -\delta)
\end{equation}
as follows.  Let $W_1,\ W_2,\ W_3,\ W_4$ be independent random variables such that 
$$\begin{array}{ll}
W_1\sim {\rm Bi}(s-1, p+\delta), &W_2\sim {\rm Bi}(s-1, p-\delta),\\
W_3\sim {\rm Bi}(t-s+1, p+\delta), &W_4\sim {\rm Bi}(t-s+1, p-\delta).
\end{array}
$$
Then 
\begin{align*}
u(s, t, \delta) &=\Pr(W_1+W_2+W_4\geq k-1);\\
u(s, t, -\delta) &=\Pr(W_1+W_2+W_3\geq k-1).
\end{align*}  
Since $\delta\geq 0$, we have
$W_4\leq_{st} W_3$, which yields $W_1+W_2+W_4\leq_{st} W_1+W_2+W_3$, and $u(s, t, \delta)\leq u(s, t, -\delta)$ by the
definition of $\leq_{st}$.  Now (\ref{ust3}), (\ref{fst}) and (\ref{uleq}) give
$$\frac{d h(\delta)}{d \delta}\leq 0$$
i.e., $h(\delta)$ decreases in $\delta$.
\end{IEEEproof}

\begin{IEEEproof}[Proof of Proposition \ref{main}]
Given the basic properties of majorization, we only need to prove that $E\phi(\sum_{i=1}^n Z_i)$ is Schur concave as a
function of $(p_1, p_2)$ holding $p_3,\ldots, p_n$ fixed.  Define $\psi(z)=E\phi(z+\sum_{i=3}^n Z_i)$.  Since $\phi$ is 
convex, so is $\psi$.  (We may assume that $\psi$ is finite as the general case can be handled by a standard limiting 
argument.) Proposition \ref{prop3}, however, shows precisely that $E\psi(Z_1+Z_2)=E\phi(\sum_{i=1}^n Z_i)$ is Schur-concave 
in $(p_1, p_2)$.
\end{IEEEproof}

\section*{Acknowledgement}
The author would like to thank O.\ Johnson and D.\ Chafa\"{i} for stimulating discussions.

\end{document}